\documentclass[submission]{eptcs}
\providecommand{\event}{PrePost17} 
\usepackage{breakurl}             
\usepackage{underscore}           

\usepackage{listings}
\usepackage{xcolor}

\lstdefinestyle{base}{
  language=Java,
  emptylines=1,
  breaklines=true,
  basicstyle=\ttfamily\color{black}\footnotesize,
  moredelim=**[is][\color{red}]{@}{@},
}

\newtheorem{definition}{Definition}
\newtheorem{example}{Example}

\usepackage{algorithmicx}
\usepackage{algorithm}
\usepackage[noend]{algpseudocode}

\usepackage{tikz}
\usetikzlibrary{positioning}
\usetikzlibrary{arrows,automata,snakes,arrows,shapes}
\tikzset{initial text={}}
\tikzstyle{accept}=[fill=gray]
\tikzstyle{skip}=[circle,draw=black]
\tikzstyle{next}=[rectangle,draw=black,minimum size=5mm]
\tikzstyle{quant}=[rectangle,dashed,draw=black,minimum size=5mm]
\newcommand{\obs}[1]{\ensuremath{\tt #1}}

\title{A Story of Parametric Trace Slicing, Garbage\\ and Static Analysis}
\author{Giles Reger
\institute{University of Manchester, Manchester, UK}
\email{giles.reger@manchester.ac.uk}
}
\def\titlerunning{A Story of Parametric Trace Slicing, Garbage and Static Analysis}
\def\authorrunning{G. Reger}
\begin{document}
\maketitle

\begin{abstract}
This paper presents a proposal (story) of how statically detecting unreachable objects (in Java) could be used to improve a particular runtime verification approach (for Java), namely parametric trace slicing. 
Monitoring algorithms for parametric trace slicing depend on garbage collection to (i) cleanup data-structures storing monitored objects, ensuring they do not become unmanageably large, and (ii) anticipate the violation of (non-safety) properties that cannot be satisfied as a monitored object can no longer appear later in the trace. The proposal is that both usages can be improved by making the unreachability of monitored objects explicit in the parametric property and statically introducing additional instrumentation points generating related events. 
The ideas presented in this paper are still exploratory and the intention is to integrate the described techniques into the MarQ monitoring tool for quantified event automata. 
%
%
%
%
\end{abstract}

\section{Introduction}

This paper explores ideas for improving the performance of runtime verification based on parametric trace slicing by the static identification of unreachable objects. 
Runtime verification \cite{Falcone2013} is the task of checking whether a given property holds on a given execution of a computational system. Typically an execution is abstracted as a \emph{trace} (e.g. a finite sequence of observations called events) and runtime verification becomes checking whether this trace is in the language defined by the property. Checking can be \emph{online} whilst the monitored program is running, or \emph{offline} after the program has run. In either case, the monitored system must be \emph{instrumented} to produce events of interest, either to be immediately processed by a monitor or to be stored in a log file.

I am particularly concerned with \emph{parametric} runtime verification for Java programs. The term parametric refers to the events being observed being annotated with parameters, such as for example $\obs{open}(\mathtt{readme.txt})$ or $\obs{insert}(\mathit{idABC},12)$. Within the context of monitoring Java programs I will assume that parametric events are generated by method calls with the event name being the method name and the parameters being a relevant subset of the method parameters and target object\footnote{This assumption is not a restriction of the technique, but convenient for this paper.}. We can, for convenience, assume that generation of such events is achieved by the Aspect-Oriented Programming (AOP) tool AspectJ. The (parametric) properties being checked are then defined in terms of these parametric events. \emph{Parametric trace slicing} is then a method for checking parametric properties that involves \emph{slicing} a parametric trace up based on parameter values and then processing each slice in separation.

\begin{figure}
\begin{lstlisting}[language=Java]
public static void writeToFile(String fileName, Collection records){
  File file = new File(fileName);
  file.open();
  Iterator iterator = records.iterator();
  while(iterator.hasNext()){
    file.write(iterator.next());
  }
  // file.close();
  records.removeAll();
}
\end{lstlisting}
\caption{Example Java program used as running example.\label{fig:java:example}}
\end{figure}

I will now use a running example (which we will use throughout the paper) to clarify the notion of parametric property, introduce the technique of parametric trace slicing, and discuss the main ideas of this paper related to the detection of unreachable objects. I keep the presentation informal here in the introduction and make things (a little) more formal later. 
Figure~\ref{fig:java:example} presents a Java method that takes a file name and a collection of records and writes the records to a file with the given name. 
I will consider three properties that I might want to check on this method.
\begin{itemize}
	\item \emph{HasNext}: For every iterator object $i$ we only call $i.\obs{next}()$ if a preceding call of $i.\obs{hasNext}()$ returned true with no intermediate calls to $i.\obs{next}()$ or $i.\obs{hasNext}()$.
	\item \emph{UnsafeIter}: For every collection $c$ and iterator object $i$ created from $c$, the iterator $i$ is not used (e.g. by calls to $i.\obs{next}()$) after $c$ has been updated.
	\item \emph{OpenClose}: For every file object $f$, the file cannot be written to or closed if not opened, cannot be opened once already open, and must eventually be closed once opened.
\end{itemize}

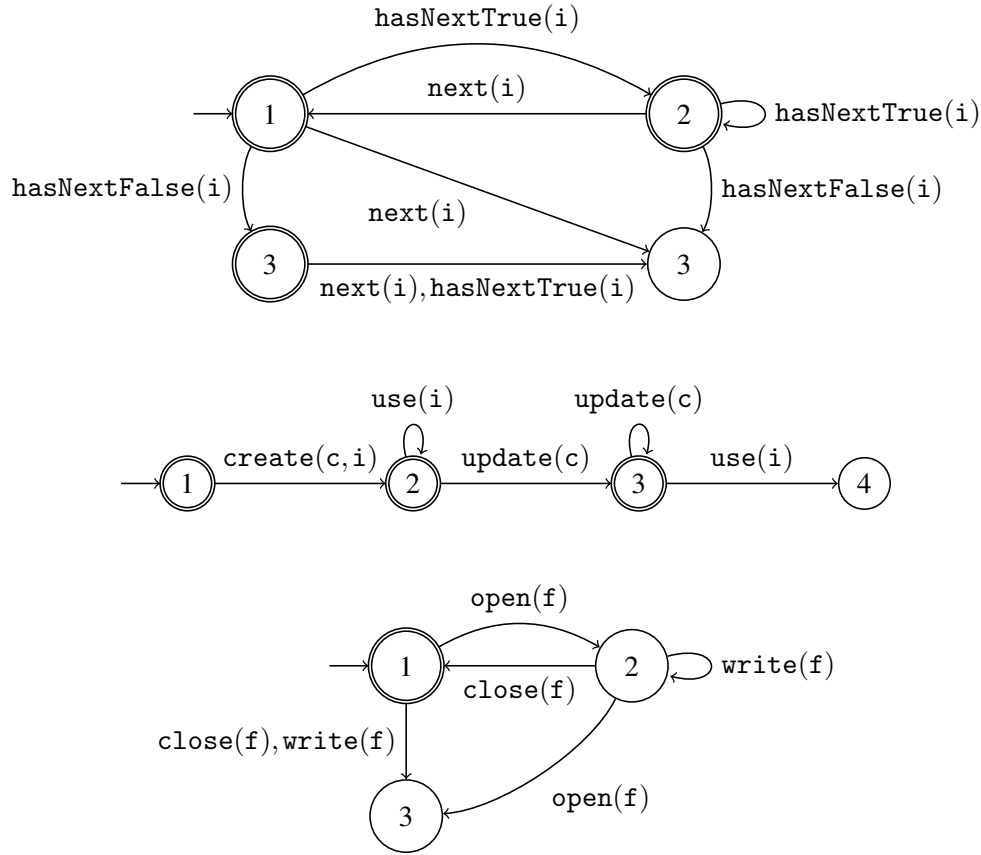
\begin{figure}[t]
\centering

\begin{tikzpicture}[->,auto,node distance=5.5cm,semithick]
\node[initial,state,accepting] (A) {1};
\node[state,accepting] (B) [right of=A] {2};
\node[state,accepting] (C) [below of=A, node distance=2cm] {3};
\node[state] (D) [below of=B, node distance=2cm] {3};
\path (A) edge  [bend left]     node {$\obs{hasNextTrue}(i)$} (B)
         (B) edge [swap]   node {$\obs{next}(i)$} (A)
         (B) edge [loop right] node {$\obs{hasNextTrue}(i)$} (B)
         (A) edge [swap] node {$\obs{next}(i)$}(D)
         (B) edge [bend left,looseness=0.7] node {$\obs{hasNextFalse}(i)$}(D)
         (A) edge [bend right,looseness=0.7,swap] node {$\obs{hasNextFalse}(i)$}(C)
         (C) edge [swap] node {$\obs{next}(i),\obs{hasNextTrue}(i)$} (D)
;
\end{tikzpicture}

\vspace{0.8cm}

\begin{tikzpicture}[->,auto,node distance=3cm,semithick]
\node[initial,skip,accepting] (A) {1};
\node[skip,accepting] (C) [right of=A] {2};
\node[skip,accepting] (D) [right of=C] {3};
\node[skip] (E) [right of=D] {4};
\path (A) edge  []   node {$\obs{create}(c,i)$} (C)
         (C) edge  []   node {$\obs{update}(c)$} (D)
         (D) edge  []   node {$\obs{use}(i)$} (E)
         (C) edge [loop above] node {$\obs{use}(i)$}(C)
         (D) edge [loop above] node {$\obs{update}(c)$}(D)
;
\end{tikzpicture}\\

\vspace{0.8cm}

\begin{tikzpicture}[->,auto,node distance=3cm,semithick]
\node[initial,state,accepting] (A) {1};
\node[state] (B) [right of=A] {2};
\node[state] (C) [below of=A, node distance=2cm] {3};
\path (A) edge  [bend left]     node {$\obs{open}(f)$} (B)
         (B) edge []   node {$\obs{close}(f)$} (A)
         (B) edge [loop right] node {$\obs{write}(f)$} (B)
         (A) edge [swap] node {$\obs{close}(f), \obs{write}(f)$}(C)
         (B) edge [bend left,looseness=0.7] node {$\obs{open}(f)$}(C)
;
\end{tikzpicture}
\caption{State machines for the three example properties.\label{fig:props}}
\end{figure}

The first two represent \emph{safety} properties whilst the third is a form of \emph{response} property as the opening of a file creates an obligation that it is eventually closed. Figure~\ref{fig:props} gives state machines for each of the properties (their semantics with respect to parameters will become clear later) where I make the assumption that states are closed to failure (to avoid an unnecessary clutter of transitions). 

Let us use the \emph{HasNext} property to illustrate the idea around parametric trace slicing. As the program is executed various Iterator objects will be created and used. Observing this will produce a \emph{trace} of such calls. For example we might observe the trace
\[
\obs{hasNextTrue}(\mathit{iter1}). \obs{next}(\mathit{iter1}).\obs{hasNextTrue}(\mathit{iter1}).\obs{hasNextTrue}(\mathit{iter2}). \obs{next}(\mathit{iter2}). \obs{next}(\mathit{iter1}).
\]
where calls to $\mathit{iter2}$ are nested within calls to $\mathit{iter1}$. The idea of parametric trace slicing is to \emph{slice} this parametric trace into two propositional traces containing event names only and check these using standard methods (e.g. using a finite state automaton). Here the trace slice for $\mathit{iter1}$ is $\obs{hasNextTrue}.\obs{next}.$ $\obs{hasNextTrue}.\obs{next}$ and for $\mathit{iter2}$ it is just $\obs{hasNextTrue}.\obs{next}$ and both slices satisfy the property.

These properties play the following role in this paper:
\begin{itemize}
	\item For the \emph{HasNext} property we can observe that the \verb|iterator| object does not escape the above \verb|writeToFile| method. This means that we could insert a call to the monitor at the end of the method indicating that the monitor no longer needs to track this object. This can be helpful to the monitoring effort as it eagerly reduces the size objects being stored by the monitor.
	\item The \emph{UnsafeIter} propery allows us to apply a similar argument within the context of another object (\verb|records|) still being potentially live.
	\item The \emph{OpenClose} property allows us to discuss \emph{anticipation}. Normally this property cannot be satisfied or violated on a finite prefix of a trace. However, it is common to consider the end of a program as indicating that no further events occur, which gives us a single infinite extension allowing us to decide satisfaction or violation. This approach therefore delays this decision to the end of the program, which has a number of issues. Now we can observe that the \verb|file| object becomes unreachable at the end of this method. A good monitoring approach would detect a violation of this property when the associated object is garbage collected, as discussed below. However, eagerly detecting this unreachability and statically adding instrumentation alerting the monitor of this, allows the monitor to detect the violation as early as possible.
\end{itemize}

\paragraph{Contributions.} Whilst this paper does not present an implementation of any of the ideas informally described above, it does present the background and ideas that form the basis for such an implementation. 
In Section~\ref{sec:detectGarbage} I review existing work on static detection of unreachable objects i.e. garbage, and in Section~\ref{sec:pts} I introduce parametric trace slicing and discuss the various ways in which information from this static analysis can help. I finish in Sections~\ref{sec:discussion} and \ref{sec:conclusion} by discussing the expected impact of these ideas and setting out my agenda for future work in this area.

\section{Statically Detecting Unreachable Objects} \label{sec:detectGarbage}

Here I briefly review various approaches for statically determining the lifetime of an object i.e. points in the code where we can be certain that an object will become unreachable. The first part discusses general techniques (these standard techniques can be found in textbooks \cite{Nielson:2010:PPA:1965094}) whilst the second part focusses on one particular approach. 

\subsection{The Intended Result}

With respect to the overall idea being put forward by this paper, the actual static analysis techniques employed do not matter. What matters is the result of the analysis and how this can be used to optimise the runtime verification approach. Therefore, I will start by discussing what the result should be.

The goal is to perform a code transformation to insert instrumentation points into the program to record when an object will no longer be observed by the monitor. This should be sound i.e. such instrumentation points should only be added where it is certain that the object can no longer be observed. I introduce this at the (informal) level of instrumentation points as this is also the level at which I introduce the runtime verification approach later.

At this point I note that in general I assume the previous goal to mean that the object becomes unreachable in general. However, a stronger analysis would take a monitored property into account and attempt to identify objects whose use becomes irrelevant to that property. Whilst this might sound like something we want, it would most likely be prohibitively expensive.

The next two sections will review some existing static analysis approaches and is not meant to be a contribution, rather a review to establish that the above goal can be met.

\subsection{Escape Analysis}

Escape analysis \cite{Blanchet:1998:EAC:268946.268949,Choi:1999:EAJ:320385.320386} is used to determine if an object escapes a method\footnote{It is also applied to check if an object escapes a thread with the goal or removing synchronisation information, but this usage is not of current interest here. Perhaps it would be interesting to explore this later for automatically `desynchronising' monitoring activities.} usually with the goal of swapping heap allocation for stack allocation as a compiler optimisation aimed at reducing memory overhead (and garbage collector load).

This analysis is typically \emph{flow-insensitive}, meaning that the order of statements in a method is ignored, and \emph{intraprocedural}, meaning that the analysis operates on a single method at a time. So let's take a look at the method of Figure~\ref{fig:java:example}. We can identify (via calls to \verb|new| and the \verb|Iterator()| factory method\footnote{Note that this requires us to explicitly identify this method as a factory method. Techniques for this are discussed later.}) two allocated objects \verb|file| and \verb|iterator| and extract the lines referring to these objects\footnote{Here I will present these ideas at the source-code level but operations could equally (and perhaps more effectively) be applied at the bytecode level.}.

\begin{center}
\begin{minipage}{0.39\textwidth}
\begin{lstlisting}[style=base]
File @file@ = new File(fileName);
@file@.open();
@file@.write(iterator.next());
\end{lstlisting}  
\end{minipage}
\hspace{0.08\textwidth}
\begin{minipage}{0.49\textwidth}  
\begin{lstlisting}[style=base]
Iterator @iterator@ = records.iterator();
while(@iterator@.hasNext())
file.write(@iterator@.next());
\end{lstlisting}    
\end{minipage}
\end{center}

It is straightforward to observe that none of these lines cause either object to escape the method. To check this programatically it is necessary to be able to identify precisely which statements cause an object to escape. There are three such kinds of statement:
\begin{enumerate}
	\item Returning the object from the method
	\item Passing the object to another method (including constructors)
	\item Creating a reference to this object from another object (which also escapes)
\end{enumerate}

Flow-insensitive analysis as described above provides an over-approximation of the escaping objects for two reasons. Firstly, it is based on the assumption that all lines in the code are reachable. Detecting unreachable parts of the code is called \emph{dead code analysis}. One could imagine combining these analyses but I suspect that unreachable code is not that common. Secondly, consider the following code where the \verb|file| variable is reassigned. The usages of the two files will be conflated in a flow-insensitive analysis, meaning that some non-escaping objects may not be identified.

\begin{lstlisting}[style=base]
...
File file = new File(fileName);
anotherMethod(file);
file = new File(anotherFileName);
...
\end{lstlisting}

A simple solution to this issue is to first place the code in \emph{static single assignment} (SSA) form which requires every variable to be assigned to exactly once and provides \emph{flow sensitivity} for local variables. 
This is a standard transformation and perhaps assuming that it will be applied is quite reasonable. 

The next (standard) challenge is that of \emph{aliasing} i.e. ensuring that all versions of a single object are tracked. For example in the following code the fact that \verb|file2| does not escape tells us that \verb|file1| does not escape as both variables \emph{point to} the same object.
\begin{lstlisting}[style=base]
...
File file1 = new File(fileName);
File file2 = file;
anotherMethod(file2);
...
\end{lstlisting}
Various \emph{pointer analysis} methods exist to determine sets of variables that necessarily or possibly point to the same heap-allocated objects. Once such sets have been computed (discussed further below) then the previous analysis is lifted to sets of aliasing variables.

Whilst this approach may identify some useful short-lived objects it has some drawbacks. Firstly, intraprocedural pointer analysis is likely to be prohibitively imprecise. Secondly, flow insensitivity is adequate for detecting escaping objects but for identifying unreachable objects early it is too conservative. Lastly, the common usage of \emph{factory methods} and \emph{pure methods} (e.g. not leaking the input parameter) is likely to significantly reduce the effectiveness of simple escape analysis for our goal. The next section introduces some existing ideas to handle these issues.

\subsection{Free-Me Analysis}

I now introduce some ideas from the \emph{free-me analysis} introduced by Guyer et al. \cite{Guyer:2006:FSA:1133255.1134024} in 2006 for identifying opportunities for early object reclamation. I am sure other equally interesting work exists (please feel free to draw my attention to it) but a discussion of this work suffices to convince us that our above goal is achievable.

\paragraph{Pointer Analysis.} The free-me analysis maintains a ${\sf points\_to}$ set for each variable (and a transitively-closed version). These sets are then modified in a flow-insensitive manner using the following rules:
\begin{center}
\begin{tabular}{lll}
Assignment & $v_1 = v_2$ & ${\sf points\_to}(v_1) ~\cup= {\sf points\_to}(v_2)$ \\
Field Access & $v_1 = v_2.f$ & ${\sf points\_to}(v_1) ~\cup= {\sf points\_to\_transitive\_closure}(v_2)$ \\
Field Assignment & $v_1.f = v_2$ & $\forall n \in {\sf points\_to}(v_1) : {\sf points\_to}(n) ~\cup={\sf points\_to}(v_2)$ \\
Static Field Access & $v_1 = Cls.f$ & ${\sf points\_to}(v_1)~\cup= {\sf points\_to}(g)$ \\
Static Field Assignment & $Cls.f = v_1$ &  ${\sf points\_to}(g) ~\cup= {\sf points\_to}(v_1)$ \\
\end{tabular}
\end{center}
A special fresh variable $g$ is used to represent all globally reachable objects and is used for global variables, static field accesses and in method summaries (discussed below). The rule for field assignment is somewhat non-standard but its conservatism allows for simple method summaries.

\paragraph{Method Summaries.} A method summary is a set of pairs $(p,q)$ where $q$ is an input parameter and $p$ is either an input parameter or the special labels $\mathit{global}$ or $\mathit{return}$. The semantics of $(p,q)$ is that after the method is called the object pointed to by $q$ is reachable from $p$. In the above pointer analysis this allows ${\sf points\_to}$ sets to be appropriately updated after method calls (for virtual methods all matching method summaries are applied). 
The analysis also identifies factory methods as those whose return object is always a newly allocated object and includes this information in the summary. In the pointer analysis factory methods are replaced by new allocation sites.


\begin{figure}
\centering
\begin{minipage}{0.5\textwidth}
\begin{lstlisting}[style=base]
File write(String value, boolean ret){
  File file = newTempFile(); // factory method
  file.write(value);
  if(ret){ return file; }
  return null;
}  
\end{lstlisting}
\end{minipage}
\hspace{0.03\textwidth}
\begin{minipage}{0.45\textwidth}
\begin{tikzpicture}
\node[draw](A){\verb|File file = newTempFile();|};
\node[draw,below=of A.west,anchor=west](B){\verb|file.write(value);|};
\node[draw,below= of B.west,anchor=west](C){\verb|if(ret)|};
\node[below= of C.west,anchor=west](D){};
\node[draw,right of=D, node distance=3cm](E){\verb|return file;|};
\node[draw,below= of D.west,anchor=west](F){\verb|return null;|};
\path[->] 
(A) edge  [out=0,in=5,pos=0.5,looseness=2,below]     node {1} (B)
(B) edge  [out=0,in=0,pos=0.5,looseness=2,below]     node {2} (C)
(C) edge  [in=180,out=270,above,pos=0.7]     node {3} (E)
(C) edge  [in=110,out=270,left]     node {4} (F)
;
\end{tikzpicture}
\end{minipage}
\caption{A small example with control flow graph.\label{fig:cfg}}
\end{figure}
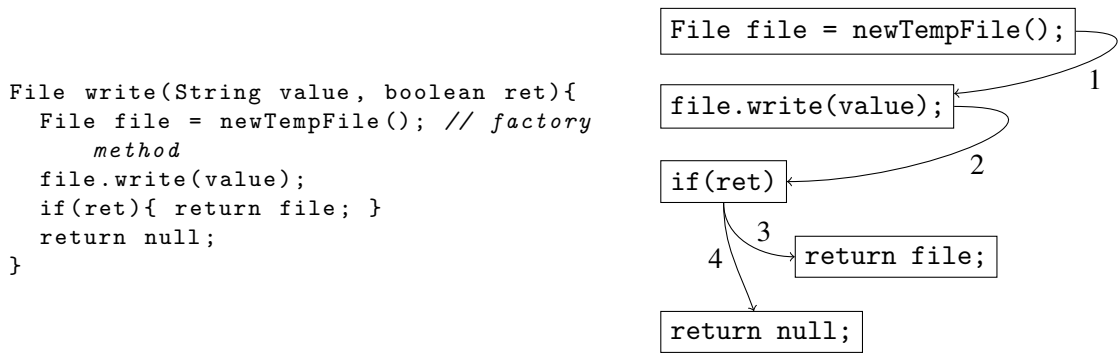

\paragraph{Liveness Analysis.} Consider the method in Figure~\ref{fig:cfg} where the \verb|file| object possibly escapes the method but there is one path through the method where it does not. Ideally we would add an instrumentation call just before the final line to indicate that the current version of the \verb|file| object will become unreachable. But the escape analysis described above will not achieve this. 

The idea is to apply traditional \emph{liveness analysis} \cite{Khedker:2009:DFA:1592955} to identify at each program point which variables contain values that may be needed later. This is applied as a backward analysis on the control flow graph where liveness is introduced at points where a variable is used and propagated backwards with unions being taken at merge points. This is applied at the granularity level of single statements (rather than basic blocks) as the usage is different from that of the traditional analysis.

This information is combined with pointer analysis to give \emph{reachability} for the objects associated with each allocation point at each edge in the control flow graph. However, the flow-insensitivity of the pointer-analysis provides an over-approximation of reachability if, for example, an object eventually becomes globally reachable. The analysis introduces a concept of \emph{potential liveness} restricting the inherited liveness status from global variables to after the associated assignment. Additionally, objects originating from factory methods are not marked as globally reachable due to the above approach of replacing such calls by allocations. 

The described approach allows us to identify that edge 4 in the small control flow graph of Figure~\ref{fig:cfg} is unreachable for the object pointed to by \verb|file|. Further details are beyond this review and I invite the reader to refer to the original text \emph{free-me analysis}.

\paragraph{Inserting Calls.}

The above analysis will identify, for each allocation site, a set of program points (edges in the control flow graph) where the object becomes unreachable. Instrumentation calls can be inserted at each site but some points will dominate others and the earliest of these should be selected. It is important to ensure that an object is not flagged as unreachable multiple times; although this is far less important in our setting than in the setting of the original work. The trick used by the original approach is to use a temporary variable that is set to \verb|null| after it is freed and a freeing method that detects and ignores values set to \verb|null|.

\section{Parametric Trace Slicing, Monitoring, and Garbage} \label{sec:pts}

Here we introduce parametric trace slicing and its relation to monitoring whilst also considering how the garbage information identifiable by the techniques in the previous section can be used to optimise monitoring. 
The monitoring approach presented here is based on the original work of Ro\c{s}u and Chen \cite{Chen2009}, which was later adapted by myself and others \cite{Barringer2012} in the definition of quantified event automata (QEA). Whilst QEA will feature at the end of this story, most of the ideas here are directly related to the original work around JavaMOP \cite{Meredith2011}.

\subsection{The Underlying Approach}

We assume disjoint sets of event names $\Sigma$, variables $X$ and values $D$. 
A valuation $\theta$ is a map (partial function with finite domain) from variables to values. The submap operator $\sqsubseteq$ and least-upper-bound operator $\sqcup$ on maps are defined as usual. We write ${\sf dom}(\theta)$ for the domain of $\theta$.

A \emph{parametric event signature} is a pair consisting of an event name $e$ and a list of variables $x_1,\ldots,x_n$ written $e(x_1,\ldots,x_n)$. A \emph{parametric event alphabet} is a finite set of parametric event signatures with at most one signature per event name. 
For a given parametric event alphabet $A$, a \emph{parametric event} is a pair consisting of an event name $e$ and a valuation $\theta$, written $e(\theta)$, such that there exists $e(x_1,\ldots,x_n) \in A$ such that $\{ x_1,\ldots,x_n \} = {\sf dom}(\theta)$. A \emph{parametric trace} is a \emph{finite} sequence of parametric events and a \emph{propositional trace} is a finite sequence of event names. Let $\epsilon$ be the empty sequence. 
A \emph{parametric property} for a parametric event alphabet $A$ is a predicate over $A^*$ (parametric traces over $A$). 
A \emph{propositional property} is a predicate over $\Sigma^*$.

We assume an instrumentation method that extracts events from programs. There is an implicit assumption that an event name corresponds to a method name and the associated variables correspond to a subset of method parameters and return value. This is not necessary but standard, particularly in the early work on parametric trace slicing. We note that assuming that a method call can only correspond to a single event in the system limits expressiveness, which is dealt with in the QEA work \cite{Barringer2012} but this solution complicates matters and we ignore it here.

For the purposes of this paper, a \emph{parametric specification} is a finite state automaton with a parametric event alphabet. Let the \emph{propositional abstraction} of this automaton be the one given by preserving event names only i.e. with alphabet $\Sigma$.

Next we will define how a parametric specification describes a parametric property. 
The idea is to use a slicing operator to transform the trace-checking problem for this parametric property into one of checking a propositional property on a set of propositional traces. The slicing operator is defined as follows.

\begin{definition}[Parametric Trace Slicing]
Given valuation $\theta$ and parametric trace $\tau$ let $\tau \downarrow_\theta$ be the propositional trace-slice for $\theta$ defined as follows.
\[
\begin{array}{lll}
\epsilon \downarrow_\theta & = & \epsilon \\
e(\theta') \tau   \downarrow_\theta& =&
\begin{array}{ll}
e (\tau   \downarrow_\theta) & ~~\mathit{if} \theta' \sqsubseteq \theta \\
(\tau   \downarrow_\theta) & ~~\mathit{otherwise} \\
\end{array}
\end{array}
\]
\end{definition}
This preserves those event names where the valuation is relevant to (included in) $\theta$. 

\begin{example}
Given a valuation $\theta = [c \mapsto A, i \mapsto B]$ and parametric trace
\[
\tau = \obs{use}(D)~\obs{create}(A,B)~\obs{create}(A,C)~\obs{use}(B)~\obs{use}(C)~\obs{update}(A)~\obs{use}(B)
\]
the corresponding trace slice for $\theta$ is
\[
\tau \downarrow_\theta = \obs{create}~\obs{use}~\obs{update}~\obs{use}
\]
\end{example}

The possible valuations used in slicing are dependent on the parametric trace being checked i.e. they are built from (parts of) those valuations observed at runtime. We define such valuations as follows.

\begin{definition}[Induced Valuations]
Given a trace $\tau$ the valuation $\theta$ is \emph{induced} by $\tau$ if (i) there is an event $e(\theta') \in \tau$ such that $\theta' \sqsubseteq \theta$, and (ii) for every $(x \mapsto v) \in \theta$ there exists an event $e(\theta') \in \tau$ such that $(x \mapsto v) \in \theta'$.
\end{definition}

\begin{example}
For the trace $\tau$ from Example 1 we have the induced valuations
\[
~[c \mapsto A, i \mapsto B] \quad [c \mapsto A, i \mapsto C] \quad [c \mapsto A, i \mapsto D]
\]
although the last valuation can be identified as \emph{redundant} using techniques not described here.
\end{example}

By construction, if a valuation $\theta$ is not induced by $\tau$ then $\tau \downarrow_\theta = \epsilon$. If the initial state of a parametric specification is accepting then this means that restricting to induced valuations does not alter the next definition but we do not enforce this restriction. 

Finally, we can define the parametric traces accepted by a parametric specification as follows.

\begin{definition}[Trace Acceptance]
A parametric specification $\Gamma$ accepts a parametric trace $\tau$ if for every valuation $\theta$ induced by $\tau$ we have
\[
\tau \downarrow_\theta \in {\cal L}(P)
\]
where ${\cal L}(P)$ is the language of the propositional abstraction of $\Gamma$.
\end{definition}

\begin{example}
Given the valuation $\theta$ and trace $\tau$ from Example 1 and the parametric specification for the \emph{UnsafeIter} proeprty given in Figure~\ref{fig:props}. We can see that the slice $\tau \downarrow_\theta$ given in Example 1 is not in the language of the propositional abstraction as it reaches state 4, which is non-final. Therefore, the trace is not accepted as there is at least one induced valuation where the given property does not hold.
\end{example}

We have now defined the semantics of our parametric specifications. Note that Figure~\ref{fig:props} already defined three parametric specifications. 

To check a trace against a parametric property then requires three steps. The first step is to construct the set of valuations, which first requires extracting values from the trace. This set is likely to be very large and theoretically exponential in the length of the trace, although usually much smaller in practice. The second step is to slice the trace to produce a slice per valuation. The final step is to check each slice against the underlying property. Clearly separating monitoring into three steps like this is not practical. 
%
The next question is how we efficiently monitor such specifications without separating monitoring into three separate steps.

\subsection{Monitoring Algorithms for Parametric Trace Slicing}

The above semantics is non-incremental due to the need to compute induced valuations before performing slicing. A number of algorithms exist for incremental monitoring \cite{Meredith2011,Reger2015}. One of the most simple of these approaches is captured in Algorithm~\ref{alg:incremental}. Here trace slices are represented by the state reached in the (propositional abstraction of the) parametric specification. The idea is to, for each incoming event, search through existing valuations and (i) update the information for the associated trace slice if the event is relevant, and (ii) add new valuations if they do not exist. Iterating through existing valuations from biggest to smallest is necessary to ensure that the valuation storing the most information about a trace slice is used when adding a new valuation (this idea is called maximality in other work).

\begin{algorithm}[t]
        \caption{An incremental algorithm for performing parametric trace slicing.\label{alg:incremental}}
        \begin{algorithmic}[1]

    \State{Let Lookup be a map from valuations to states initial mapping the empty valuation to the initial state}
    \For{event $e(\theta) \in \tau$}
      \For{$\theta'$ in ${\sf dom}$(Lookup) from biggest to smallest}
      \If{$\theta$ is consistent with $\theta'$}
        \If{$\theta' \sqsubseteq \theta$}
        		\State{ Update Lookup($\theta'$) using $e$}
	\ElsIf{$\theta \sqcup \theta'$ is not in ${\sf dom}$(Lookup)}
		\State{Add $\theta \sqcup \theta'$ to Lookup using Lookup($\theta'$) updated using $e$}
        \EndIf
      \EndIf
      \EndFor
    \EndFor
    \State If an entry in Lookup is in a non accepting state then Fail otherwise Accept
        \end{algorithmic}
      \end{algorithm}

I note two things. Firstly, that this algorithm is heavily dependent on the size of Lookup, which in the given algorithm never shrinks. And secondly, the algorithm is inefficient as each step is linear in the number of stored valuations, and in the worse case the number of valuations can be exponential in the length of the trace seen so far e.g. $v^{|\tau|}$ where $v$ is the number of variables used in the parametric specification. Although this assumes heavy reuse of values in the trace and typically the number of valuations does not grow exponentially, but may still grow super-linearly.

Efficient monitoring approaches relying on complex indexing data structures have been introduced (see \cite{Meredith2011,Reger2015}) but these remain super-linearly related to the number of valuations as they employ redundancy to ensure efficient indexing. Whilst this is often a suitable trade-off, there is still a cost associated with maintaining such structures and access operations remain somewhat proportional to the size of the structures. 
The message here is that \emph{the number of valuations being tracked directly impacts efficiency}. We Therefore, reducing this number (for example by detecting that objects in a valuation become unreachable and relevant events can no longer be observed) will have a positive effect.

\subsection{Anticipation}

The reader may have noticed a problem with Algorithm~\ref{alg:incremental}. Even though the trace is processed incrementally the result of whether the full trace is accepted only comes at the end. It is possible to report on the acceptance of the current prefix (i.e. what would happen if the trace ended here) but this does not necessarily relate to the final verdict. 

What we want to do is \emph{anticipate} the final verdict as soon as possible. 
For the \emph{HasNext} and \emph{UnsafeIter} properties given in Figure~\ref{fig:props} this is relatively straightforward as they are safety-properties. As soon as any entry in Lookup enters a non-accepting state there is no way for the trace to be accepted and final violation can be reported straight away. For the \emph{OpenClose} property the safety element can be detected early but the eventually closing a file part cannot (as it stands).

What is the formal idea here? If the current trace cannot be \emph{extended} to an accepting trace then it can be marked as non-accepting. We can capture the notion of possible extensions by \emph{reachability} in the finite state automaton. In the \emph{HasNext} finite state automata failure is captured by the explicit failure state 3 where, whereas in the \emph{UnsafeIter} automata there is (in addition to the explicit state 4) an implicit failure state where, by construction, no accepting states are reachable. In the \emph{OpenClose} automata state 3 has no reachable accepting states so any entries in Lookup entering this state can report a final violation.

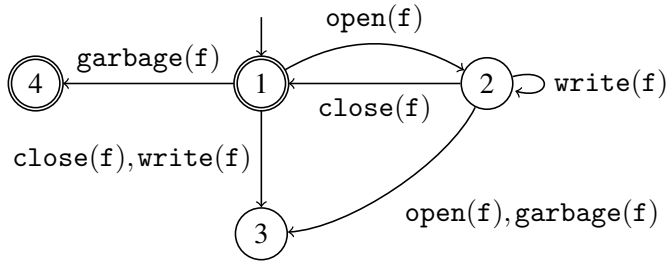
\begin{figure}[t]
\centering
\begin{tikzpicture}[->,auto,node distance=3cm,semithick, initial where=above]
\node[initial,skip,accepting] (A) {1};
\node[skip] (B) [right of=A] {2};
\node[skip] (C) [below of=A, node distance=2cm] {3};
\node[skip,accepting] (D) [left of = A] {4};
\path (A) edge  [bend left]     node {$\obs{open}(f)$} (B)
         (B) edge []   node {$\obs{close}(f)$} (A)
         (B) edge [loop right] node {$\obs{write}(f)$} (B)
         (A) edge [swap] node {$\obs{close}(f), \obs{write}(f)$}(C)
         (B) edge [bend left,looseness=0.7] node {$\obs{open}(f), \obs{garbage}(f)$}(C)
         (A) edge [swap] node {$\obs{garbage}(f)$} (D)
;
\end{tikzpicture}
\caption{Adding $\obs{garbage}$ events to the \emph{OpenClose} automaton.\label{fig:garbage:openclose}.}
\end{figure}

\begin{figure}[t]
\begin{tikzpicture}[->,auto,node distance=3cm,semithick]
\node[initial,skip,accepting] (A) {1};
\node[skip,accepting] (C) [right of=A] {2};
\node[skip,accepting] (D) [right of=C] {3};
\node[skip,accepting] (C2) [below of=C, node distance=2cm] {2c};
\node[skip,accepting] (C3) [above of=C, node distance=2cm]{2b};
\node[skip,accepting] (D3) [right of=C3] {3b};
\node[skip] (E) [right of=D] {5};
\path (A) edge  []   node {$\obs{create}(c,i)$} (C)
         (C) edge  []   node {$\obs{update}(c)$} (D)
         (D) edge  []   node {$\obs{use}(i)$} (E)
         (C) edge [in=310,out=350,looseness=8] node {$\obs{use}(i)$}(C)
         (D) edge [loop below] node {$\obs{update}(c)$}(D)
         (C2) edge [loop below] node {$\obs{use}(i)$}(C2)     
         (C) edge [swap] node {$\obs{garbage}(c)$}(C2)           
         (C) edge [pos=0.8] node {$\obs{garbage}(i)$}(C3)         
         (D) edge [pos=0.8] node {$\obs{garbage}(i)$}(D3)                  
         (C3) edge  []   node {$\obs{update}(c)$} (D3)
         (D3) edge [loop right] node {$\obs{update}(c)$}(D3)         
;
\end{tikzpicture}
\caption{Adding $\obs{garbage}$ events to the UnsafeIter property.\label{fig:garbage:UnsafeIter}}
\end{figure}
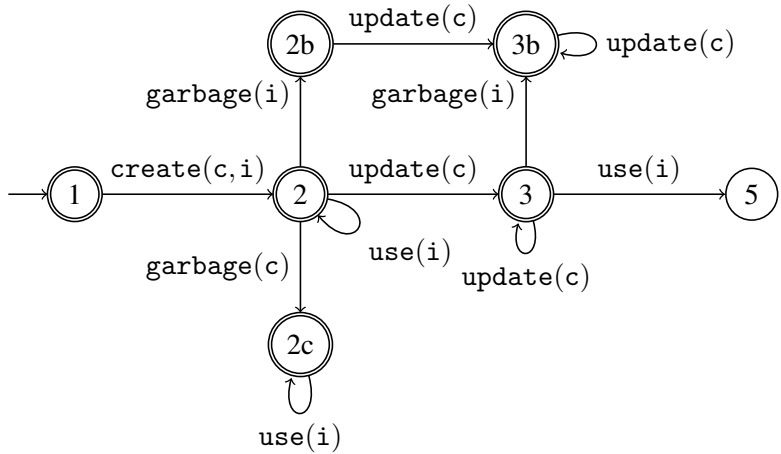

This is the first point where detection of unreachable objects can help. If, for the \emph{OpenClose} property we can detect that a file object becomes unreachable when the associated automaton is in the non-accepting state 2 then we know that this cannot be remedied later. 
Instead of adding special code to the monitoring approach to handle this case\footnote{Although I note that it is standard to have special monitoring code, along with the usage of \emph{weak references}, to handle such cases, as discussed later.} we can simply modify the automaton for the property and add a $\obs{garbage}$ event (produced by the techniques discussed earlier) to the alphabet of the property. Figures~\ref{fig:garbage:openclose} and ~\ref{fig:garbage:UnsafeIter} show how this can be done for the \emph{OpenClose} and \emph{UnsafeIter} properties respectively. Importantly, the addition of garbage events can be an automatic transformation applied to automata. 

An observation here is that for the \emph{UnsafeIter} property if $c$ becomes garbage at state 3 it is still necessary to continue monitoring $i$ as a violation can still occur, however if this occurs earlier then the property cannot be violated any longer. Furthermore, detecting that $i$ becomes garbage means that the property can never be violated as  the only paths to a non-accepting state involve the $i$ object. Therefore, the valuation can be removed if $i$ becomes garbage (which is the case in our example from Figure~\ref{fig:java:example}) but not necessarily if $c$ becomes garbage.

\subsection{Garbage-Aware Indexing}

Indexing structures employed by the monitoring algorithms mentioned above are typically \emph{garbage-aware} in the sense that they employ \emph{weak references} so that once all objects pointed to by a valuation become unreachable that entry in the data structure can be removed. The initial implementation of this idea in JavaMOP was incorrect as it did not allow for the anticipation of failure as described above (which led to missing some property violations), this was fixed later \cite{jin-meredith-griffith-rosu-2011-pldi}. 

The usage of weak references is important as it prevents the monitor \emph{leaking} memory and it keeps data structures small. However, doing this via garbage collection has a disadvantage as objects are not collected until space is needed.
Therefore, the second usage of static garbage information would be to eagerly reduce the size of these data structures by directly reporting this garbage. 

However, I note that this may have limited impact as the garbage issues are larger for longer-living objects whilst the static analysis techniques proposed earlier are more suitable for short-lived objects. Nevertheless, in situations where many short-lived objects are being monitored at the same time I see this as potentially having a positive impact.

\subsection{Monitoring Representative Objects}

This idea is the least mature but also has the potential for having the most impact on monitoring overhead. My observation is that when we perform static analysis we consider all objects produced by a single allocation point together. In runtime verification we do not do this; every created object is monitored separately. This makes sense for long-lived objects with many possible paths through the program but for short-lived objects it is a large waste of monitoring time.

For example, in the method given in Figure~\ref{fig:java:example} we should only monitor \verb|file| and \verb|iterator| once although this becomes less clear in the presence of the loop. If this method is called frequently then only monitoring the method once will have a massive impact on overhead.

The proposal is to detect objects that do not escape a method and add flags to indicate whether that object has been checked on all paths through the method. The reason that escape analysis is a key factor here is that we want to be able track all paths of a single object and doing this interprocedurally would be prohibitive. 
For our example method it is straightforward as there is a single path. For methods with complex control flow this may be prohibitively complicated. In essence, the proposal is to monitor an \emph{abstraction} of observed concrete objects and whilst the idea here is to use an allocation-site abstraction, this could be further augmented for greater precision.

A related approach was previously proposed by Dwyer et al. in 2010 \cite{Dwyer2010}. The idea there was to optimise loops to reduce monitoring overhead by detecting repeated iterations that did not differ in their effect, unrolling them, and only monitoring the first iteration. It appears that Dwyer et al.'s idea is related to this one and but has not been fully explored.

\subsection{Offline Monitoring}

A small but sometimes important point is that in implementations of parametric trace slicing the notion of equality is typically referential. This is reasonable for online monitoring as at any one time only one object can occupy a particular memory address. However, for offline monitoring where events are recorded in a log file and then read in later for checking there needs to be a way of unambiguously recording object identities. Simply printing the identity hash code is insufficient as there is the possibly of a clash with an object created later. Whilst this sounds improbable, I have experienced this in practice.

The solution is to either add allocation or garbage information to the log file. Allocation information is not always appropriate as we may only be monitoring a subset of a certain type of object that is used in a particular way and logging all allocations may increase the size of the trace by orders of magnitude. 

Currently, I use a trick that creates objects with a custom \verb|finalizer| weakly reachable by a monitored object such that collection of this object records unreachability of the monitored object. However this relies on finalizers, which are not guaranteed to be run, and requires additional objects. Therefore, static detection and reporting of unreachability is preferable (although sadly not a general solution).

\section{Discussion or Will it Work?} \label{sec:discussion}

The previous section proposed a number of approaches for using static information about unreachable objects to optimise the runtime verification effort. Here I discuss what impact I might expect these to have in general. Before doing so it is worth mentioning that tools such as JavaMOP and MarQ depend heavily on garbage information at runtime and without ideas such as garbage-aware indexing they would struggle with a real-world trace of any significant length. However, I note that whilst the impact of garbage has been studied \cite{jin-meredith-griffith-rosu-2011-pldi}, the previous view of representing garbage within the property being monitored is new.

\paragraph{Risks and Limitations.}
The main limitation of this approach is the scope of applicability. The static analyses discussed earlier are only applicable to objects that are are short lived and (relatively) local to a single method. Arguably, in such cases it could be relatively easy to apply static techniques to completely verify a property. Indeed, the more interesting properties for runtime verification are those involving objects and events spread out over time and the codebase. Due to realistic language features (such as reflection and dynamic loading) it is very difficult to make the described static analyses sound inter-procedurally i.e. many objects will be conservatively marked as reachable rendering the approach ineffective for non-local properties. Whilst more modern approaches \cite{Smaragdakis:2015:PA:2802194.2802195,DBLP:conf/ecoop/SpathDAB16} may provide better precision, this will remain a limitation of the approach (however, see the other side of this below).

A related point to the above is that for such short-lived objects it is highly likely that modern generational garbage collection will collect objects quickly, providing the desired quick anticipation of liveness violations. The current use of weak references generally appears adequate and there is a risk that any advantage provided by static information could be represent a small increment at best. 



\paragraph{Perceived Strengths.} A counter to the above point about short-lived objects is that this approach can make use of partial information (where an object is unreachable in some paths only) in situations where intraprocedural static checking would not apply. Additionally, for properties involving multiple objects, this approach may only identify that one object becomes unreachable, which would not be usable information statically, but could have significant impact dynamically.

I expect the idea about monitoring representative objects to be the most fruitful in the long-term. Whilst the other ideas may decrease overhead per monitored object, this approach has the potential to remove the need to monitor a large number of objects at all.


\section{Conclusion} \label{sec:conclusion}

This paper has explored the idea of statically identifying unreachable objects and then using this information to optimise runtime verification using parametric trace slicing by
\begin{itemize}
	\item anticipating failures of non-safety properties sooner than otherwise possible,
	\item keeping indexing data structures small,
	\item reducing the number of objects being monitored (in restricted circumstances), and
	\item dealing with a known issue with offline monitoring in this setting.
\end{itemize}
None of these ideas have been implemented yet but the intention is to incorporate them into the MarQ \cite{Reger2015} tool. I note that supporting the additional features of QEA beyond those captured by the parametric specifications described in this paper may require some extra work. For example, QEA also allow objects to be captured by so-called \emph{free variables} that are not involved in slicing and then for transitions to be guarded by predicates on these variables. This dramatically increases the complexity of the reachability question as it must consider the satisfiability of these guards.

As a final comment, this work began in \cite{Reger2016} by exploring how the typestate analysis techniques employed by Clara \cite{Bodden2010, DBLP:journals/sttt/BoddenH12} could be applied to QEA. The idea was to utilise pointer analysis information to statically check non-safety properties. However, I then noticed that the same information could be used to optimise the runtime activity in different ways. This work looking at typestate analysis for QEA is also ongoing.

\paragraph{Acknowledgements.} I would like to thank the reviewers for their helpful comments that helped improve the text and also provide further ideas to explore.

\bibliographystyle{eptcs}
\bibliography{bib}

\begin{thebibliography}{10}
\providecommand{\bibitemdeclare}[2]{}
\providecommand{\surnamestart}{}
\providecommand{\surnameend}{}
\providecommand{\urlprefix}{Available at }
\providecommand{\url}[1]{\texttt{#1}}
\providecommand{\href}[2]{\texttt{#2}}
\providecommand{\urlalt}[2]{\href{#1}{#2}}
\providecommand{\doi}[1]{doi:\urlalt{http://dx.doi.org/#1}{#1}}
\providecommand{\bibinfo}[2]{#2}

\bibitemdeclare{inproceedings}{Barringer2012}
\bibitem{Barringer2012}
\bibinfo{author}{Howard \surnamestart Barringer\surnameend},
  \bibinfo{author}{Yli{\`e}s \surnamestart Falcone\surnameend},
  \bibinfo{author}{Klaus \surnamestart Havelund\surnameend},
  \bibinfo{author}{Giles \surnamestart Reger\surnameend} \&
  \bibinfo{author}{David~E. \surnamestart Rydeheard\surnameend}
  (\bibinfo{year}{2012}): \emph{\bibinfo{title}{Quantified Event Automata:
  Towards Expressive and Efficient Runtime Monitors}}.
\newblock In: {\sl \bibinfo{booktitle}{FM}}, pp. \bibinfo{pages}{68--84}.
\newblock \urlprefix\url{http://dx.doi.org/10.1007/978-3-642-32759-9_9}.

\bibitemdeclare{inproceedings}{Blanchet:1998:EAC:268946.268949}
\bibitem{Blanchet:1998:EAC:268946.268949}
\bibinfo{author}{Bruno \surnamestart Blanchet\surnameend}
  (\bibinfo{year}{1998}): \emph{\bibinfo{title}{Escape Analysis: Correctness
  Proof, Implementation and Experimental Results}}.
\newblock In: {\sl \bibinfo{booktitle}{Proceedings of the 25th ACM
  SIGPLAN-SIGACT Symposium on Principles of Programming Languages}},
  \bibinfo{series}{POPL '98}, \bibinfo{publisher}{ACM}, \bibinfo{address}{New
  York, NY, USA}, pp. \bibinfo{pages}{25--37}, \doi{10.1145/268946.268949}.

\bibitemdeclare{article}{DBLP:journals/sttt/BoddenH12}
\bibitem{DBLP:journals/sttt/BoddenH12}
\bibinfo{author}{Eric \surnamestart Bodden\surnameend} \&
  \bibinfo{author}{Laurie~J. \surnamestart Hendren\surnameend}
  (\bibinfo{year}{2012}): \emph{\bibinfo{title}{The Clara framework for hybrid
  typestate analysis}}.
\newblock {\sl \bibinfo{journal}{{STTT}}}
  \bibinfo{volume}{14}(\bibinfo{number}{3}), pp. \bibinfo{pages}{307--326},
  \doi{10.1007/s10009-010-0183-5}.
\newblock \urlprefix\url{https://doi.org/10.1007/s10009-010-0183-5}.

\bibitemdeclare{inproceedings}{Bodden2010}
\bibitem{Bodden2010}
\bibinfo{author}{Eric \surnamestart Bodden\surnameend},
  \bibinfo{author}{Patrick \surnamestart Lam\surnameend} \&
  \bibinfo{author}{Laurie \surnamestart Hendren\surnameend}
  (\bibinfo{year}{2010}): \emph{\bibinfo{title}{Clara: A Framework for
  Partially Evaluating Finite-state Runtime Monitors Ahead of Time}}.
\newblock In: {\sl \bibinfo{booktitle}{Proceedings of the First International
  Conference on Runtime Verification}}, \bibinfo{series}{RV'10},
  \bibinfo{publisher}{Springer-Verlag}, \bibinfo{address}{Berlin, Heidelberg},
  pp. \bibinfo{pages}{183--197}, \doi{10.1007/978-3-642-16612-9_15}.

\bibitemdeclare{inproceedings}{Chen2009}
\bibitem{Chen2009}
\bibinfo{author}{Feng \surnamestart Chen\surnameend} \&
  \bibinfo{author}{Grigore \surnamestart Ro\c{s}u\surnameend}
  (\bibinfo{year}{2009}): \emph{\bibinfo{title}{Parametric Trace Slicing and
  Monitoring}}.
\newblock In: {\sl \bibinfo{booktitle}{TACAS '09}}, \bibinfo{address}{Berlin,
  Heidelberg}, pp. \bibinfo{pages}{246--261},
  \doi{10.1007/978-3-642-00768-2_23}.

\bibitemdeclare{article}{Choi:1999:EAJ:320385.320386}
\bibitem{Choi:1999:EAJ:320385.320386}
\bibinfo{author}{Jong-Deok \surnamestart Choi\surnameend},
  \bibinfo{author}{Manish \surnamestart Gupta\surnameend},
  \bibinfo{author}{Mauricio \surnamestart Serrano\surnameend},
  \bibinfo{author}{Vugranam~C. \surnamestart Sreedhar\surnameend} \&
  \bibinfo{author}{Sam \surnamestart Midkiff\surnameend}
  (\bibinfo{year}{1999}): \emph{\bibinfo{title}{Escape Analysis for Java}}.
\newblock {\sl \bibinfo{journal}{SIGPLAN Not.}}
  \bibinfo{volume}{34}(\bibinfo{number}{10}), pp. \bibinfo{pages}{1--19},
  \doi{10.1145/320385.320386}.

\bibitemdeclare{inbook}{Dwyer2010}
\bibitem{Dwyer2010}
\bibinfo{author}{Matthew~B. \surnamestart Dwyer\surnameend},
  \bibinfo{author}{Rahul \surnamestart Purandare\surnameend} \&
  \bibinfo{author}{Suzette \surnamestart Person\surnameend}
  (\bibinfo{year}{2010}): \emph{\bibinfo{title}{Runtime Verification in
  Context: Can Optimizing Error Detection Improve Fault Diagnosis?}}, pp.
  \bibinfo{pages}{36--50}.
\newblock \bibinfo{publisher}{Springer Berlin Heidelberg},
  \bibinfo{address}{Berlin, Heidelberg}, \doi{10.1007/978-3-642-16612-9_4}.

\bibitemdeclare{incollection}{Falcone2013}
\bibitem{Falcone2013}
\bibinfo{author}{Y.~\surnamestart Falcone\surnameend},
  \bibinfo{author}{K.~\surnamestart Havelund\surnameend} \&
  \bibinfo{author}{G.~\surnamestart Reger\surnameend} (\bibinfo{year}{2013}):
  \emph{\bibinfo{title}{A Tutorial on Runtime Verification}}.
\newblock In \bibinfo{editor}{Manfred \surnamestart Broy\surnameend} \&
  \bibinfo{editor}{Doron \surnamestart Peled\surnameend}, editors: {\sl
  \bibinfo{booktitle}{Summer School Marktoberdorf 2012 - Engineering Dependable
  Software Systems}}, \bibinfo{publisher}{IOS Press},
  \doi{10.3233/978-1-61499-207-3-141}.

\bibitemdeclare{article}{Guyer:2006:FSA:1133255.1134024}
\bibitem{Guyer:2006:FSA:1133255.1134024}
\bibinfo{author}{Samuel~Z. \surnamestart Guyer\surnameend},
  \bibinfo{author}{Kathryn~S. \surnamestart McKinley\surnameend} \&
  \bibinfo{author}{Daniel \surnamestart Frampton\surnameend}
  (\bibinfo{year}{2006}): \emph{\bibinfo{title}{Free-Me: A Static Analysis for
  Automatic Individual Object Reclamation}}.
\newblock {\sl \bibinfo{journal}{SIGPLAN Not.}}
  \bibinfo{volume}{41}(\bibinfo{number}{6}), pp. \bibinfo{pages}{364--375},
  \doi{10.1145/1133255.1134024}.

\bibitemdeclare{inproceedings}{jin-meredith-griffith-rosu-2011-pldi}
\bibitem{jin-meredith-griffith-rosu-2011-pldi}
\bibinfo{author}{Dongyun \surnamestart Jin\surnameend},
  \bibinfo{author}{Patrick~O'Neil \surnamestart Meredith\surnameend},
  \bibinfo{author}{Dennis \surnamestart Griffith\surnameend} \&
  \bibinfo{author}{Grigore \surnamestart Ro\c{s}u\surnameend}
  (\bibinfo{year}{2011}): \emph{\bibinfo{title}{Garbage Collection for
  Monitoring Parametric Properties}}.
\newblock In: {\sl \bibinfo{booktitle}{Proceedings of the 32nd ACM SIGPLAN
  conference on Programming Language Design and Implementation (PLDI'11)}},
  \bibinfo{publisher}{ACM}, pp. \bibinfo{pages}{415--424},
  \doi{10.1145/1993498.1993547}.

\bibitemdeclare{book}{Khedker:2009:DFA:1592955}
\bibitem{Khedker:2009:DFA:1592955}
\bibinfo{author}{Uday \surnamestart Khedker\surnameend},
  \bibinfo{author}{Amitabha \surnamestart Sanyal\surnameend} \&
  \bibinfo{author}{Bageshri \surnamestart Karkare\surnameend}
  (\bibinfo{year}{2009}): \emph{\bibinfo{title}{Data Flow Analysis: Theory and
  Practice}}, \bibinfo{edition}{1st} edition.
\newblock \bibinfo{publisher}{CRC Press, Inc.}, \bibinfo{address}{Boca Raton,
  FL, USA}, \doi{10.1201/9780849332517}.

\bibitemdeclare{article}{Meredith2011}
\bibitem{Meredith2011}
\bibinfo{author}{Patrick \surnamestart Meredith\surnameend},
  \bibinfo{author}{Dongyun \surnamestart Jin\surnameend},
  \bibinfo{author}{Dennis \surnamestart Griffith\surnameend},
  \bibinfo{author}{Feng \surnamestart Chen\surnameend} \&
  \bibinfo{author}{Grigore \surnamestart Ro{\c s}u\surnameend}
  (\bibinfo{year}{2011}): \emph{\bibinfo{title}{An overview of the MOP runtime
  verification framework}}.
\newblock {\sl \bibinfo{journal}{J Software Tools for Technology Transfer}},
  pp. \bibinfo{pages}{1--41}.
\newblock \urlprefix\url{http://dx.doi.org/10.1007/s10009-011-0198-6}.

\bibitemdeclare{book}{Nielson:2010:PPA:1965094}
\bibitem{Nielson:2010:PPA:1965094}
\bibinfo{author}{Flemming \surnamestart Nielson\surnameend},
  \bibinfo{author}{Hanne~R. \surnamestart Nielson\surnameend} \&
  \bibinfo{author}{Chris \surnamestart Hankin\surnameend}
  (\bibinfo{year}{2010}): \emph{\bibinfo{title}{Principles of Program
  Analysis}}.
\newblock \bibinfo{publisher}{Springer Publishing Company, Incorporated}.

\bibitemdeclare{inproceedings}{Reger2016}
\bibitem{Reger2016}
\bibinfo{author}{Giles \surnamestart Reger\surnameend} (\bibinfo{year}{2016}):
  \emph{\bibinfo{title}{Considering Typestate Verification for Quantified Event
  Automata}}.
\newblock In: {\sl \bibinfo{booktitle}{7th International Symposium on
  Leveraging Applications of Formal Methods, Verification and Validation (ISoLA
  2016)}}, \doi{10.1007/978-3-319-23820-3_21}.

\bibitemdeclare{inproceedings}{Reger2015}
\bibitem{Reger2015}
\bibinfo{author}{Giles \surnamestart Reger\surnameend},
  \bibinfo{author}{Helena~Cuenca \surnamestart Cruz\surnameend} \&
  \bibinfo{author}{David \surnamestart Rydeheard\surnameend}
  (\bibinfo{year}{2015}): \emph{\bibinfo{title}{{MarQ}: monitoring at runtime
  with {QEA}}}.
\newblock In: {\sl \bibinfo{booktitle}{Proceedings of the 21st International
  Conference on Tools and Algorithms for the Construction and Analysis of
  Systems (TACAS'15)}}, \doi{10.1007/978-3-662-46681-0_55}.

\bibitemdeclare{article}{Smaragdakis:2015:PA:2802194.2802195}
\bibitem{Smaragdakis:2015:PA:2802194.2802195}
\bibinfo{author}{Yannis \surnamestart Smaragdakis\surnameend} \&
  \bibinfo{author}{George \surnamestart Balatsouras\surnameend}
  (\bibinfo{year}{2015}): \emph{\bibinfo{title}{Pointer Analysis}}.
\newblock {\sl \bibinfo{journal}{Found. Trends Program. Lang.}}
  \bibinfo{volume}{2}(\bibinfo{number}{1}), pp. \bibinfo{pages}{1--69},
  \doi{10.1561/2500000014}.

\bibitemdeclare{inproceedings}{DBLP:conf/ecoop/SpathDAB16}
\bibitem{DBLP:conf/ecoop/SpathDAB16}
\bibinfo{author}{Johannes \surnamestart Sp{\"{a}}th\surnameend},
  \bibinfo{author}{Lisa Nguyen~Quang \surnamestart Do\surnameend},
  \bibinfo{author}{Karim \surnamestart Ali\surnameend} \& \bibinfo{author}{Eric
  \surnamestart Bodden\surnameend} (\bibinfo{year}{2016}):
  \emph{\bibinfo{title}{Boomerang: Demand-Driven Flow- and Context-Sensitive
  Pointer Analysis for Java}}.
\newblock In: {\sl \bibinfo{booktitle}{30th European Conference on
  Object-Oriented Programming, {ECOOP} 2016, July 18-22, 2016, Rome, Italy}},
  pp. \bibinfo{pages}{22:1--22:26}, \doi{10.4230/LIPIcs.ECOOP.2016.22}.
\newblock \urlprefix\url{https://doi.org/10.4230/LIPIcs.ECOOP.2016.22}.

\end{thebibliography}
\end{document}